\documentclass[conference]{IEEEtran}
\usepackage{graphicx}
\usepackage[caption=false,font=footnotesize]{subfig}
\usepackage{cite}

\usepackage[normalem]{ulem}
\usepackage{amsmath}
\usepackage{bbm}
\usepackage[caption=false]{subfig}
\usepackage{tikz}
\usepackage{xcolor}
\usetikzlibrary{quantikz2}
\usepackage{float}
\usepackage{algorithm}
\usepackage{algpseudocode}
\usepackage{amsfonts}

\definecolor{color1bg}{HTML}{676DC5}
\usepackage[colorlinks=true,urlcolor=blue, citecolor=blue, linkcolor=blue, plainpages=false]{hyperref}  
\usepackage{pgfplots}
\usepgfplotslibrary{groupplots}
\pgfplotsset{compat=1.18}
\pgfplotsset{                   
    /pgfplots/legend image code/.code={%
    \draw[mark repeat=2,mark phase=2,#1]                
        plot coordinates {(0cm,0cm) (0.3cm,0cm)};%
        }
    }
        
\definecolor{cbBlue}{HTML}{0072B2}    
\definecolor{cbOrange}{HTML}{E69F00}  
\definecolor{cbGreen}{HTML}{009E73}   
\definecolor{cbRed}{HTML}{D55E00}     
\definecolor{cbPurple}{HTML}{CC79A7}  
\definecolor{cbYellow}{HTML}{F0E442}  
\definecolor{cbSky}{HTML}{56B4E9}     

\allowdisplaybreaks

\title{Quantum Lattice Boltzmann Solutions for Transport
under 3D Spatially Varying Advection on Trapped‑Ion Hardware}

\author{
\IEEEauthorblockN{
Sayonee Ray\IEEEauthorrefmark{1}\textsuperscript{\textsection},
Jezer Jojo\IEEEauthorrefmark{2}\textsuperscript{\textsection},
Jason Iaconis\IEEEauthorrefmark{1},
Abeynaya Gnanasekaran\IEEEauthorrefmark{1},
Apurva Tiwari\IEEEauthorrefmark{2}\\
Martin Roetteler\IEEEauthorrefmark{1},
Chris Hill\IEEEauthorrefmark{2},
Jay Pathak\IEEEauthorrefmark{2}
}\medskip
\IEEEauthorblockA{\IEEEauthorrefmark{1}IonQ Inc., 4505 Campus Dr., College Park, MD 20740, U.S.A.}
\IEEEauthorblockA{\IEEEauthorrefmark{2}Synopsys Inc., U.S.A.}
}
\date{}

\begin{document}

\maketitle
\begingroup\renewcommand\thefootnote{\textsection}
\footnotetext{Equal contribution}
\endgroup

\begin{abstract}
The Quantum Lattice Boltzmann Method (QLBM) has emerged as one of the most promising quantum computing approaches for the numerical simulation of problems in computational fluid dynamics (CFD). The dynamics is formulated in terms of mesoscopic particle distribution functions governed by a discrete Boltzmann transport equation, comprising local streaming and collision operations. In this work, the resulting macroscopic behavior corresponds to the advection–diffusion equation, which we adopt as a canonical model problem for transport phenomena. Building upon recent progress in QLBM implementations, we advance towards more realistic problem settings that better reflect conventional CFD requirements. We address, for the first time, transport under the action of non-uniform velocity fields on quantum hardware. We implement our demonstration using IonQ's trapped-ion systems including Forte generation systems and a 64-qubit Barium development system similar to the forthcoming IonQ Tempo line.  We identify the density readout and subsequent reloading of the fluid density as a potential bottleneck of the current algorithm and discuss several approaches to mitigate this bottleneck. We identify the use of MPS shadow tomography as a promising method to efficiently scale the readout to large system with complex density distributions.  Lastly, we introduce and simulate a novel method to implement wall boundaries for advection-diffusion in QLBM, and discuss the prospects of scaling to higher-complexity problems.
\end{abstract}


\section{Introduction}


\begin{figure*}[t]
    \centering
    \includegraphics[width=0.90\linewidth]{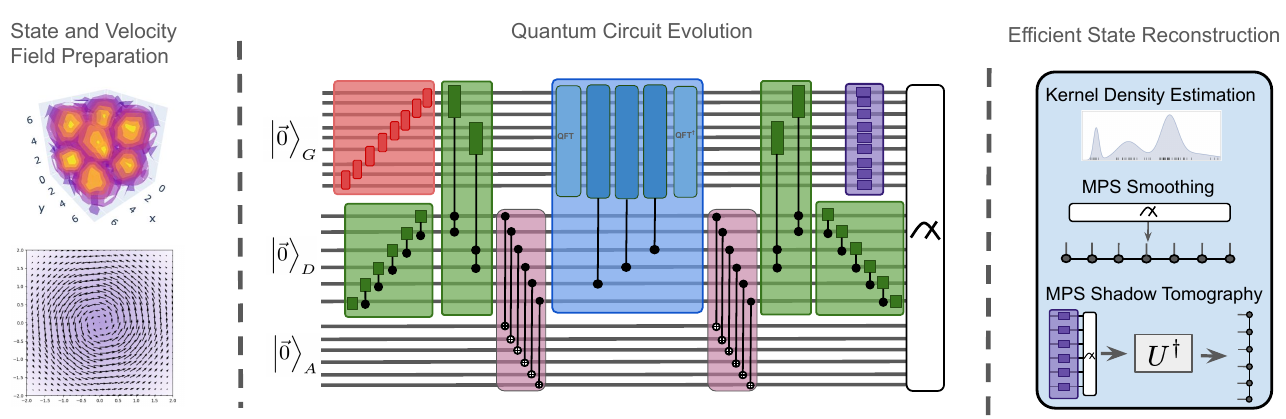}
    \caption{ QLBM pipeline used in this work. We efficiently encode the initial density distribution and velocity field into quantum operators. The MPS state preparation, streaming and collision operations are then performed within the quantum circuit. An error detection gadget using ancilla qubits is also applied. The output density is reconstructed using three potential methods described in the text: kernel density estimation, MPS smoothing, and MPS shadow tomography. }
    \label{fig: qlbm_circuit}
\end{figure*}

Partial differential equations (PDEs) are among the most important industrial targets for quantum computing. Of the many engineering domains where PDEs are used to model physical phenomena, computational fluid dynamics (CFD) \cite{LeVeque2002}, in particular, constitutes a substantial fraction of the total high‑performance computing workload, and accelerating CFD solvers remains an active and long‑standing objective within the CFD community. This has motivated growing interest in quantum algorithms aimed at accelerating CFD simulations; either through the acceleration of fundamental algorithmic building blocks, such as large‑scale ordinary differential equation (ODE) solvers arising from semi‑discretized PDEs, or through fundamentally new quantum formulations that seek to represent the PDE solution, and operate in ways enabled by the distinct primitives of quantum computation~\cite{steijl_parallel_2018}. 

A wide variety of quantum approaches to linear and nonlinear PDEs have been proposed, including quantum linear algebra solvers \cite{leyton2008quantum,arrazola2019quantum}, variational algorithms \cite{lubasch2020variational,sarma2024quantum}, and Hamiltonian simulation techniques \cite{meng2023quantum}. Within this landscape, methods inspired by the lattice Boltzmann method (LBM) \cite{mohamad_lbm} are particularly attractive for quantum implementation \cite{ljubomir2022quantum,ionq-ansys-qlbm-2025, budinskiadvec,li2025potential,wang2025quantum}. By formulating continuum dynamics through the evolution of mesoscopic particle distribution functions on discrete lattices, the LBM yields local and highly structured update rules. This structure naturally supports quantum encodings of particle densities on large lattices, and implementation of the collision and streaming operations using modest resources which scale only logarithmically with the number of grid points, making LBM‑based formulations a promising avenue for quantum CFD solvers.

In QLBM, as in the classical LBM, the dynamics are decomposed into a linear streaming step, which propagates particle distributions along a discrete set of directions, and a local nonlinear collision step, which relaxes the distribution toward equilibrium. Quantum implementations of these streaming and collision operators have been developed in prior work for fluid-flow simulations \cite{dineshqlbm,budinskiadvec}. Early formulations, however, faced several practical bottlenecks, including large compiled gate counts, costly state preparation and readout, and vanishing post-selection success probabilities in the collision step. Recent algorithmic advances have substantially reduced these limitations~\cite{ionq-ansys-qlbm-2025}, providing a framework which was used to simulate advection-diffusion of a Gaussian wavepacket in two dimensions under a constant advection field. In this work, we build on that foundation in several ways. We demonstrate a QLBM simulation of three-dimensional advection-diffusion in a spatially varying swirl field on trapped-ion hardware, and introduce MPS-based readout methods, including shadow-MPS tomography, that make iterative readout and state reloading practical on current devices. We also show how more general spatially varying velocity fields can be implemented efficiently within the same framework.

The readout stage presents a central challenge for near-term QLBM demonstrations. In principle, the full evolution of the particle distribution over $T$ time steps could be executed within a single circuit. In practice, however, the resulting circuit depth, combined with post-selection overhead due to hardware noise, makes this approach impractical on current devices. Periodic readout of the density distribution is also desirable for other reasons: in particular, it may enable intermediate-term implementations of nonlinear dynamics without requiring a fully dense Carleman-linearized collision operator \cite{itani2022analysis,sanavio2024three,sanavio2025explicit}. For the complex three-dimensional, spatially varying dynamics studied here, simple parameterizations of the density like those used in Ref.~\cite{ionq-ansys-qlbm-2025}, are no longer adequate. Instead, we develop new strategies for reconstructing the density distribution from a limited number of measurements. In particular, we show that matrix product states (MPS) offer an effective reconstruction and reload mechanism for the QLBM states considered here. 

To realize this idea in practice, we combine kernel density estimation (KDE) with fixed-bond-dimension MPS reconstruction. KDE smooths statistical fluctuations in the measured histograms, while the MPS model provides a compact representation of the density distribution that is effective in our QPU experiments. To scale to larger system sizes, we further introduce a shadow-MPS tomography protocol, which uses randomized single-qubit measurements to learn the MPS representation directly from measurement data. We find that this shadow-based approach outperforms direct histogram-based reconstruction at larger system sizes, where standard readout methods become increasingly measurement-limited.

In this work, we address many of the outstanding questions of how to properly scale the QLBM algorithm, with an implementation on trapped-ion quantum hardware of significantly greater complexity than previous demonstrations. The key contributions of our paper are:
\begin{itemize}
    \item \textbf{3D spatially varying QLBM on hardware:} We demonstrate advection-diffusion under a three-dimensional swirl transporting field on a trapped-ion quantum computer.
    \item \textbf{Efficient readout and reloading:} We show that KDE-assisted MPS reconstruction enables accurate iterative readout and re-preparation of QLBM states on hardware.
    \item \textbf{Scalable shadow-MPS tomography:} We introduce a classical shadow-based MPS tomography method that improves reconstruction fidelity at larger system sizes and lower effective shot counts.
    \item \textbf{Extending to wall boundaries and more general flows:} We further show in simulation the framework to implement nontrivial velocity fields and wall boundary conditions with improved efficiency over earlier methods.
\end{itemize}

The rest of this paper is organized as follows. In Section~\ref{sec:intro}, we describe the model we will focus on in this work and outline how to construct quantum circuits to simulate the dynamics using the QLBM. In Section \ref{sec:hw_demo}, we describe the main experimental results demonstrated on the IonQ Tempo-line Barium system. We further discuss inherent noise resilience and the application of different readout strategies on our results.  In Section~\ref{sec:shadows}, we apply a new approach based on classical shadows combined with MPS tomography and demonstrate superior readout as the system size scales with this method. In Section~\ref{sec:arbitrary_fields}, we demonstrate how our methods can be applied to efficiently simulate general spatially varying velocity fields, and in Section~\ref{sec:wall_boundaries} we implement wall boundaries, improving on prior implementations. We then present simulation results on these cases.

\subsection{Circuit construction}
For the scope of this work, we are interested in simulating advection-diffusion under a divergence-free velocity field, using the Lattice Boltzmann Method with D2Q5 and D3Q7 models and with relaxation time $\tau=1$.

Given an initial scalar field $\Phi(\vec{r},t)$, with lattice vector $\vec{r}$, and a velocity field denoted by $\vec{u}(\vec{r})$, then under the Bhatnagar-Gross-Krook (BGK) approximation \cite{BGK},

\begin{eqnarray}
    \Phi(\vec{r},t+\Delta t) &=&\sum_i k_i(\vec{r})\Phi\left(\vec{r}-\vec{c}_i\Delta t, t\right), \\
    k_i(\vec{r})&=&\omega_i\left[1+\frac{\vec{c}_i\cdot \vec{u}(\vec{r})}{c_s^2}\right],
    \label{eq:k_define}
\end{eqnarray}
where $c_i = (-1)^{i+1}\vec{e}_{\lfloor (i+1)/2 \rfloor}$ and $c_0=\vec{0}$ with $\vec{e}_i$ the $i^{\text{th}}$ basis vector. We take $c_s=1/\sqrt3$, and $\Delta t=1$. The lattice weights $\{\omega_i\}_i$ depend on the Lattice Boltzmann model.

We follow the quantum algorithm framework proposed in \cite{budinskiadvec} and modified in \cite{ionq-ansys-qlbm-2025}. The algorithm uses two main registers - the `grid' register (denoted $G$) and the `direction' register (denoted $D$) with $n$ qubits for DmQn.

We make use of the following operators. First we have the streaming operator

\begin{equation}
    U_S:=\prod_i \left[\mathbbm{1}_G\otimes (\mathbbm{1}-|i_H\rangle\langle i_H|)_D+S_i\otimes(|i_H\rangle\langle i_H|)_D\right],\label{eq:Ustream}
\end{equation}
where $S_i|x\rangle_G=|x+\vec{c}_i\rangle_G$, and $|i_H\rangle$ denotes the bitstring with a $1$ at the $i$-th position from the right and $0$ elsewhere.

Then we have the PREP and UNPREP operators denoted $U_P$ and $U_Q$ for the collision step. We define $U_P$ to be an operator that satisfies

\begin{equation}
    U_P\ket{\vec r}\ket{0} = \ket{\vec{r}}\sum_{i} \sqrt{k_i(\vec{r})}\ket{i_H}_D. \label{eq:Up}
\end{equation}
We define $U_Q$ to be an operator that satisfies
\begin{equation}
    U_Q^\dagger \ket{\vec{r}}\ket{0} =  \sum_{i} \sqrt{k_i(\vec{r}-\vec{c}_i)}\ket{i_H}) \label{eq:Uq}
\end{equation}


We encode our initial state $\Phi(\vec{r},0)$ as a statevector $|\Phi_0\rangle_G$ on the grid register. This is done using the MPS circuit construction of \cite{iaconis2023tensornetworkbasedefficient,iaconis2024quantum, holmes2020efficient,rudolph2024decomposition}. We then repeatedly apply the following circuit operations to get from $|\Phi_t\rangle$ to $|\Phi_{t+1}\rangle$:

\begin{itemize}
    \item Apply PREP as defined in Eq. \eqref{eq:Up} to get
\begin{equation}
    \frac{1}{\|\Phi_t\|}\sum_x\sum_i\sqrt{k_i(x)}\Phi(x,t)|x\rangle_G|i\rangle_D
\end{equation}
    \item Apply the streaming operator $U_S$ of Eq. \eqref{eq:Ustream}.
    \begin{align}
        &\frac{1}{\|\Phi_t\|}\sum_x\sum_i\sqrt{k_i(x)}\Phi(x,t)|x+\vec{c}_i\rangle_G|i_H\rangle_D\nonumber\\
        =&\frac{1}{\|\Phi_t\|}\sum_{x,i}\sqrt{k_i(x-\vec{c}_i)}\Phi(x-\vec{c_i},t)|x\rangle_G|i_H\rangle_D
    \end{align}
    \item Apply the UNPREP operator in Eq. \eqref{eq:Uq} to get
    \begin{equation}
        \frac{\|\Phi_{t+1}\|}{\|\Phi_t\|}|\Phi_{t+1}\rangle_G|0\rangle_D+|\xi\rangle
    \end{equation}
    where $(\mathbbm{1}_G\otimes\langle0|_D)|\xi\rangle=0$.
    \item Measure the direction qubits and post-select for the $|0\rangle$ state to yield the desired state $|\Phi_{t+1}\rangle_G$.
\end{itemize}

The state preparation, collision and streaming operators are shown in Fig.~\ref{fig: qlbm_circuit} in red, green and blue blocks respectively. We further apply and error detection gadget (shown in the pink block) using ancilla qubits and a potential shadow tomography rotation on the grid qubits (purple block). Note that this algorithm can be applied according to the readout and reload protocol in every time step or after multiple time steps. Although the majority of the results we discuss in this paper follow the first strategy, we have also explored the multiple step strategy and its accuracy on the QPU. The advantage in the latter is that we reduce the error source from the approximate tomography methods and state reloading at every step. However, we also recover fewer shots on post-selection due to hardware noise with multiple LCU cycles in each circuit execution.

 \label{sec:intro}

\section{Hardware demonstration of a 3D swirl}\label{sec:hw_demo}

\subsection{Trapped Ion Hardware}

We implemented the QLBM algorithm on IonQ's trapped-ion hardware to simulate a three-dimensional swirl velocity field in the $D3Q7$ model. On a grid of $N = 8 \times 8 \times 8$ lattice sites, we ran the algorithm for six time steps, using $50{,}000$ $Z$-basis measurement shots per step and reconstructing the state with kernel density estimation and matrix product state (MPS) fitting. This procedure reproduced the dynamics over all six time steps with fidelity relative to the exact density state, $|\langle \phi | \psi \rangle|^2$, exceeding $88\%$. As discussed below, scaling to larger grids rapidly increases the shot cost of direct reconstruction. To address this, we employ a classical-shadow-based MPS tomography protocol, which enables simulation on a grid of size $N = 16^3$ with similar fidelity using only $20{,}000$ shots measured across multiple bases.

We ran our quantum circuits on IonQ's Forte-class systems, based on Yb$^+$ ions, and on a larger IonQ-developed Barium prototype system analogous to the upcoming IonQ Tempo product line. The latter platform supports up to 64 Barium qubits arranged in a long-chain, steered-beam configuration~\cite{ionq-kipu-protein-folding}. On all of these devices, including the Forte platforms, ions are generated by laser ablation and selective ionization and are trapped in compact, integrated vacuum packages based on surface linear Paul traps. Two-photon Raman transitions driven by 355 nm laser pulses for the Ytterbium systems and 532 nm laser pulses for the Barium system provide universal gate control through arbitrary single-qubit rotations and entangling ZZ gates.

These QPUs employ sophisticated optical control systems built around acousto-optic deflectors (AODs), which substantially reduce beam alignment errors by enabling precise, independent beam steering to individual ions~\cite{Kim-aod,PRXQuantum.2.020343}. The Barium development system additionally incorporates leakage checks to identify and discard samples in which the quantum state is affected by interactions with the surrounding environment. Together, these architectural features sustain consistently high gate fidelities and make these systems particularly well suited to experiments involving deep circuits~\cite{Chen2024benchmarkingtrapped}.

We now present hardware results on the IonQ Barium system, verify that state preparation and readout can be carried out using MPS representations of the density function, and discuss the intrinsic noise resilience of the implementation, which supports circuits with $\sim300$ two-qubit gates while achieving per-time-step fidelities of up to $97\%$.

\subsection{QPU Results on  Barium System}

\begin{figure}[htbp]
    \centering
    \includegraphics[width=0.89\linewidth]{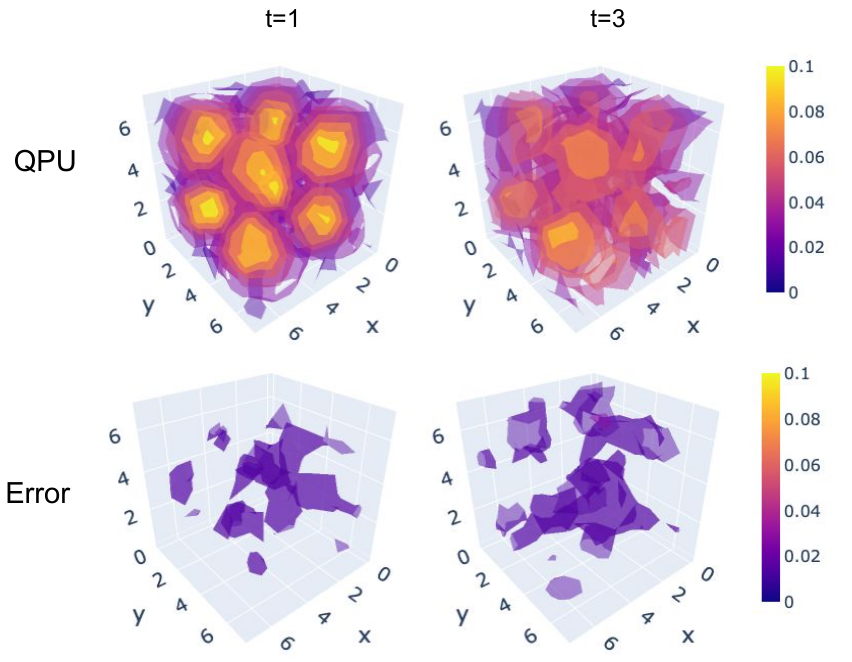}  \\
    \vspace{1pc}
   \includegraphics[width=0.88\linewidth]{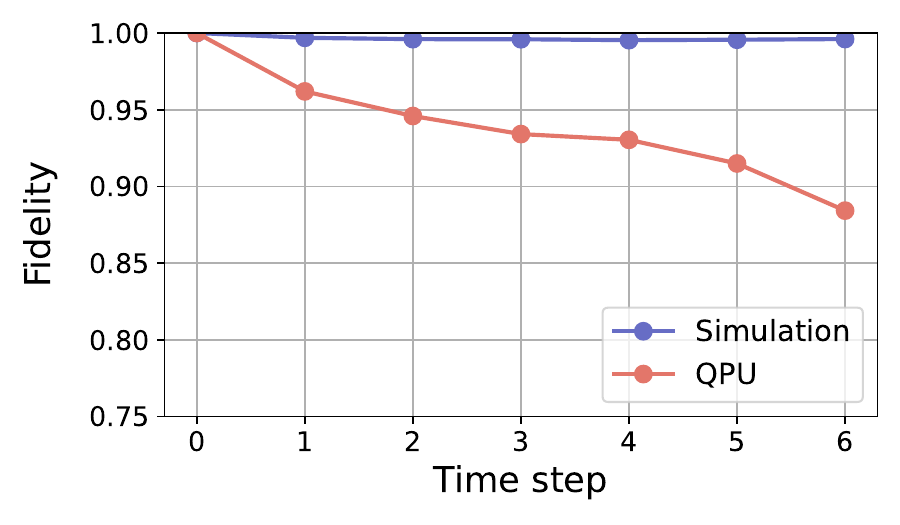}
   
    \caption{{\it Top:} Reconstructed 3D density distribution from the QPU at $t=1$ and $t=3$, together with the absolute error relative to the exact solution. {\it Bottom:} Fidelity, $|\langle \phi | \psi \rangle|^2$, for the ideal simulation and QPU runs. The simulation includes only finite-shot error, whereas the QPU data also include hardware noise.}
    \label{fig:qpu_results_1}
\end{figure} 

We simulate the $D3Q7$ swirl dynamics using the full QLBM workflow on the Tempo-line Barium system, with KDE and MPS smoothing used for readout post-processing. The circuit uses $\sim 300$ two-qubit gates on $21$ qubits, including MPS state preparation, the streaming and LCU-based collision operators, and flag-qubit error detection. At each time step, the grid-register state $\ket{\psi}_G$ is measured, post-processed, and reloaded via the MPS state-preparation circuit. Figure~\ref{fig:qpu_results_1} shows the reconstructed three-dimensional density distribution and the absolute error relative to the exact simulation at $t=1$ and $t=3$. The fidelity, $F = |\langle \psi | \phi \rangle|^2$, remains above $0.90$ for most of the evolution and reaches $F=0.88$ after six time steps.

After a single time step, the fidelity is $\sim 0.97$. To understand why the fidelity remains so high despite the circuit depth, we can examine the effect of the one-hot encoding and LCU post-selection structure. In this picture, the UNPREP operator acts as
\begin{equation}
    \ket{\psi} = \ket{0}\ket{\psi_0} + \sum_{i \in H[1]} \ket{i} \otimes \ket{\psi_i}
    \xrightarrow{\text{UNPREP}}
    \ket{0}\left(\sum_{i=0}^{Q} \ket{\psi_i}\right) + \ket{G},
\end{equation}
where $H[1]$ denotes the set of Hamming-weight-1 bitstrings. Bit-flip errors that move the direction register into the $H[2]$ sector are naturally mapped into the orthogonal subspace $\ket{G}$, preventing them from corrupting the density dynamics on the grid register. Bit-flip errors on the grid qubits can still occur without being flagged, but the data indicates that these errors have a smaller effect on the final measured state. This is reflected in the QPU histograms: in the ideal simulation, $\sim 80\%$ of shots lie in the target subspace with $\ket{0}_D$, whereas in the hardware data only $12\text{--}15\%$ do so and also pass the flag-qubit check. About $65\%$ of shots are instead mapped into the $\ket{G}$ subspace and discarded. This robustness is enabled by the one-hot encoding of the direction qubits introduced in Ref.~\cite{ionq-ansys-qlbm-2025}.



\subsection{MPS approximation of the density distributions}

While the principal advantage of QLBM is its ability to efficiently evolve density distributions on very large grids, extracting and reconstructing these distributions without resorting to exponentially costly full-state tomography remains a significant challenge. This difficulty is pronounced in our implementation, where the distribution must be measured and reconstructed at every time step. Even with $50{,}000$ measurement shots, post-selection onto the target subspace leaves only about $7500$ usable samples for reconstructing the state over an $N = 8^3$ grid, leading to substantial statistical fluctuations. We mitigate this problem by exploiting the spatial smoothness of the density function throughout the evolution: measurements from neighboring grid points are combined to infer the most likely distribution across the full grid. To do so, we employ two complementary techniques, kernel density estimation (KDE) and MPS smoothing. Applied together, these methods improve the reconstructed-state fidelity on hardware by $2$--$4\%$, which is significant in this regime.

\begin{figure}[htbp]
    \centering
    {\includegraphics[width=\linewidth,height=5cm]{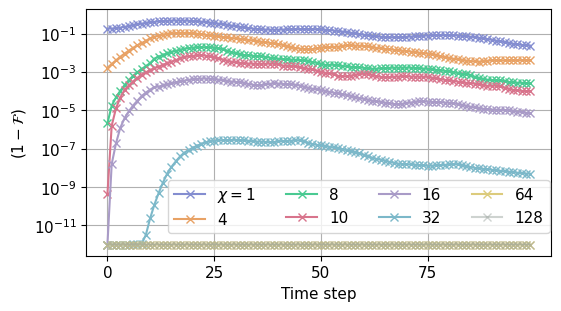}}~\\
    {\includegraphics[width=\linewidth,height=5cm]{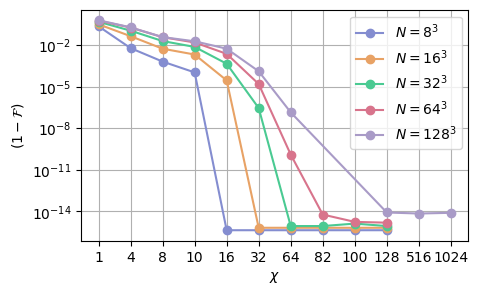}}
    \caption{(Top) Infidelity of MPS compression of ideal statevector during 3D swirl evolution for a $32 \times 32 \times 32$ grid. (Bottom) Infidelity vs bond dimension at different grid sizes at times where the infidelity of the respective MPS approximation is highest during the evolution. Large grid density distributions were determined using exact classical simulations. }
    \label{fig:mps-representation-fidelity}
\end{figure}

Indeed, MPS representations are used both in the state preparation and readout portions of our workflow, as a complex parameterization of the density distribution which can be efficiently measured. In Fig.~\ref{fig:mps-representation-fidelity}, we show that for the $N=32^3$ grid, the infidelity of the fixed bond dimension MPS representation peaks at an intermediate time, implying that the full time evolution can be effectively captured with bond dimension $\chi$. We further show that the maximum infidelity for each choice of $\chi$ collapses onto a single curve as we increase the grid size $N$. Together these results show that the full advection-diffusion dynamics of the 3D swirl flow can be accurately represented with MPS of fixed bond dimension. Note that although the state encoded on the grid qubits is well described by an MPS, the full state also includes entanglement with the direction qubits.

An efficient MPS parameterization need not exist for all fluid-dynamics problems, as fine spatial structure in the density field can produce highly entangled wavefunctions. Even so, we expect the present methods to remain useful in more complex settings. In particular, they can still be applied in the presence of an entanglement barrier~\cite{rath2023entanglement}, where the quantum circuit is used to evolve the state through a highly entangled regime, while MPS-based readout and reloading are performed on either side of the barrier.


\subsection{Tomography}
As mentioned, we implement three classical postprocessing methods: 1) Kernel density estimation (KDE), which is a standardized method of kernel smoothing for estimating probability density function from samples~\cite{kde_rosenblatt_1956, kde_parzen_1962}, 2)  smoothing based on MPS approximations of the state read out from the circuit histogram \cite{ionq-ansys-qlbm-2025, iaconis2023tensornetworkbasedefficient}, and 3) MPS shadow tomography, which uses random single qubit Pauli measurements to train a tensor network model to learn the state \cite{kuzmin2024learning}.

\subsubsection{KDE smoothing}
If $x = \{x_1, x_2, x_3, .., x_s \}$ are independent and identically distributed samples drawn from a distribution with an unknown density \textit{f}, then the kernel density estimator is:
$f_h (x) = \frac{1}{s h}\sum^s_{i=1} K\left(\frac{x-x_i}{h}\right)$
where, $K$ is the kernel and $h$ is the bandwidth or the smoothing parameter. The choice of bandwidth depends on the samples available and the underlying distribution~\cite{kde-silverman-1998}. A small bandwidth leads to less smoothing in the density estimation and vice versa. We choose the kernel to be Gaussian and the bandwidth to be $0.5$ for $50{,}000$ shots on the QPU.


\subsubsection{MPS smoothing}
Instead of reconstructing the state directly from the histogram, we fit an MPS representation of the state with fixed bond dimension. For an appropriate choice of bond dimension, this can give a improved state reconstruction.  This method when integrated with KDE further gives a high fidelity tomography technique, especially, when implemented on the QPU output. 

\begin{figure}[htbp]
    \centering
    {\includegraphics[width=0.9\linewidth]{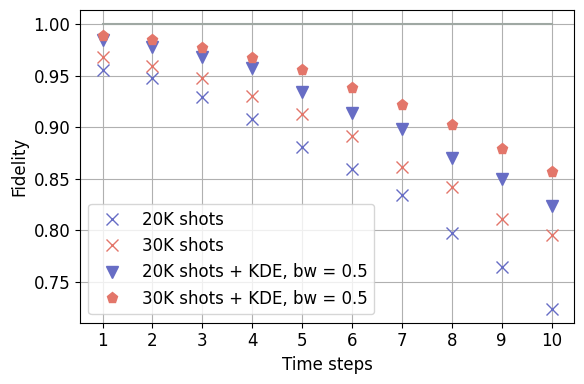}}\\
    {\includegraphics[width=0.9\linewidth]{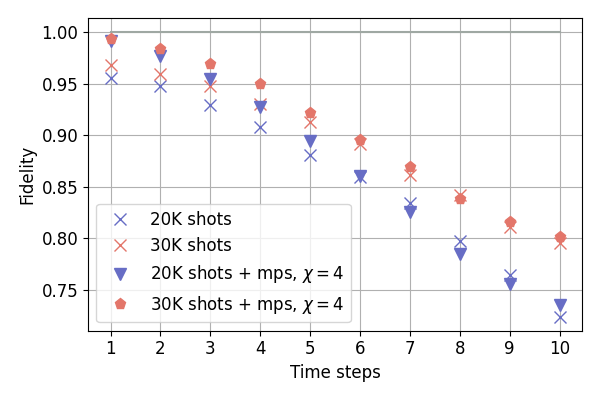}} 
    
    \caption{Fidelity $|\langle \psi|\phi\rangle|^2$ of ideal simulation with KDE (top) and MPS smoothing (bottom) techniques with statevector simulation on a $16^3$ grid. }
    \label{fig:kde_mps_smoothing}
\end{figure}

\begin{figure}[htbp]
    \includegraphics[width=1.0\linewidth]{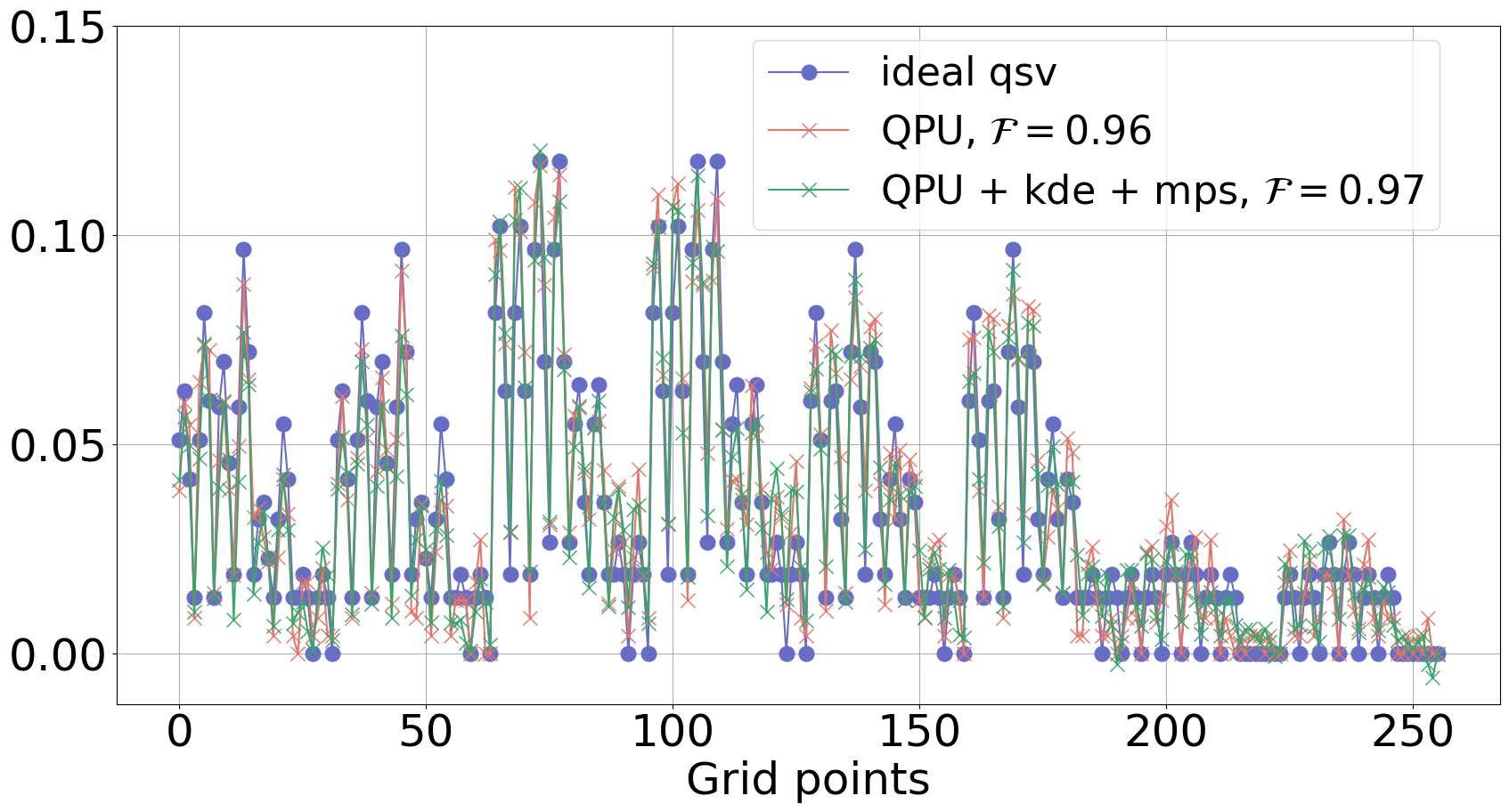} 
    \caption{Effect of KDE and MPS smoothing techniques on a $8^3$ state after $1$ time step on the Tempo class Barium system, visualized as a 1D function here. We plot only the first half for better visualization.}
    \label{fig:kde_mps_hardware}
\end{figure}

In Fig.~\ref{fig:kde_mps_smoothing}, we see that both KDE and MPS smoothing provide a noticeable improvement in the reconstructed state fidelity in simulation over the naive reconstruction. In Fig.~\ref{fig:kde_mps_hardware}, we show that applying both methods successively on the quantum hardware provides an additional improvement in fidelity, particularly improving the raw QPU (Tempo line) fidelity from $0.96$ to $0.97$.

When applying the readout-reload protocol after multiple steps, we noticed a decrease in fidelity in the raw QPU output as expected. The post-selection recovers $< 2,000$ shots out of $50{,}000$ shots. On Forte, reading out the $8^3$ grid state after $2$ time steps give a fidelity of $\sim 0.81$ with raw QPU output and gets improved to $\sim 0.94-0.95$ with the KDE and MPS smoothing methods.


\section{Scaling with MPS Shadow Tomography}\label{sec:shadows}

As shown in the previous section, the QLBM states considered here are well approximated by fixed-bond-dimension MPS. Our initial readout strategy estimates the computational-basis distribution from measurement histograms and then fits a MPS to this state. While effective at small system sizes, this approach becomes increasingly sample-inefficient as the number of grid qubits grows, since the histogram becomes sparse and the shot budget required to maintain fidelity rises rapidly. Instead we develop an alternative approach to directly learn the MPS representation from measurement data, sidestepping full state reconstruction. To aid in this, we utilize newly developed techniques based on classical shadows \cite{huang2020predicting}.

Shadow tomography provides an efficient framework for learning properties of quantum states from few randomized measurements. 
We build on the method of Kuzmin et al.~\cite{kuzmin2024learning}, which learns tensor-network states from factorized single-qubit measurements, and adapt it with a few key changes to the QLBM setting.

 A set of $M$ measurement settings is generated, each consisting of independent random \texttt{SU(2)} rotations for the $n_G = \log_2 N$ position qubits. In each setting, the single qubit unitaries are appended to the QLBM circuit in Fig.~\ref{fig: qlbm_circuit} before the resulting state is measured in the computational basis. 
A fixed number of shots $K$ is used per setting to generate an empirical probability distribution $\hat{p}_{m}(b) = n_{m}(b)/ K$ over the bitstrings, with $n_m(b)$ the observed shots per bitstring. 
Given the $M$ measurement settings and observed probability distributions, we fit a MPS $\ket{\psi}$ with bond-dimension $\chi$ by minimizing the Hellinger loss function, 
$
\mathcal{L} = \sum_{m=1}^{M} \sum_b \left(\sqrt{\hat{p}_{m}(b)}-\sqrt{p_m(b)} \right)^2.
$
Here, $p_m(b) = |\langle b | U^{(m)}_1U_2^{(m)}\dots U_{n_G}^{(m)} \ket{\psi}|^2$ is the predicted probability for bitstring $b$ under the measurement setting $m$. The \texttt{QTensor} library~\cite{qtensor} is used to optimize the MPS parameters using stochastic gradient descent (SGD).

We depart from Ref.~\cite{kuzmin2024learning} in three ways. First, we fit a pure-state MPS rather than a matrix product operator (MPO). This is appropriate because post-selection and flag-qubit filtering remove a substantial fraction of faulty shots, and because our target is the amplitude distribution rather than a full density matrix. It also reduces the number of variational parameters. Second, because many raw shots are discarded by post-selection, the one-shot-per-setting regime of Ref.~\cite{kuzmin2024learning} is inefficient in our case. We instead use fewer settings with more shots per setting, which produces better-conditioned empirical histograms and a more stable optimization signal. Third, we replace the mean-squared-error loss with Hellinger loss, which is better suited to sparse probability distributions and empirically yields higher reconstruction fidelity.

Shadow tomography is often most useful when computational-basis measurements are insufficient to determine the target state. However, the QLBM setting considered here is special: the reconstructed state encodes a fluid density field and is therefore real and non-negative in the computational basis. In principle, computational-basis measurements alone are sufficient to reconstruct the target amplitudes. However, as shown by both our simulation and hardware results in Section~\ref{sec:results_shadow}, we find that shadow-MPS tomography yields much higher-fidelity reconstructions. Here we provide intuition for this behavior.
\begin{itemize}
    \item \textit{Complementary projections of the same state.} With a finite shot budget, computational-basis measurements produce a noisy histogram with poor statistics in the tails of the distribution, and low-probability bins may receive zero counts. Measuring in $M$ randomly rotated bases provides complementary views of the same state. Outcomes that are poorly sampled in the computational basis can contribute more strongly in other bases, allowing the shadow-MPS procedure to aggregate information across multiple projections and reconstruct the state more accurately.

    \item \textit{Separation of pure-state signal from incoherent noise.} A computational-basis histogram cannot distinguish between a pure state and a mixed state that produce the same amplitudes in that basis. By contrast, rotated-basis measurements reveal information that is sensitive to incoherent noise. For example, the pure state $\ket{\psi} = a\ket{0} + b\ket{1}$ and the mixed state $\rho = |a|^2 \ket{0}\bra{0} + |b|^2 \ket{1}\bra{1}$ have identical $Z$-basis statistics but differ in the $X$ basis. When fitting a pure-state MPS to histograms collected across many rotated bases, the optimizer is forced to find a state that best explains all measurements simultaneously, which tends to suppress incoherent noise in the fitted model. Coherent errors, however, are generally more difficult to mitigate in this way.

    \item \textit{Fitting a compact model to overconstrained data.} 
     We fit a low-bond-dimension MPS with a number of variational parameters scaling as $2\,n_G\chi^2$. For typical parameters ($n_G \sim 9,18$ and $\chi \sim 3$), this corresponds to only $\sim 162$--$324$ parameters, which are fit to thousands of observed probabilities across all measurement settings. The resulting overconstrained optimization favors a smooth, self-consistent solution and helps filter shot noise and hardware noise. In contrast, in the direct-histogram approach the noise is already embedded in the reconstructed histogram before MPS compression is performed.
\end{itemize}




\subsection{Results from Forte/simulator with MPS shadows}
\label{sec:results_shadow}

\begin{figure}[!htbp]
    \centering
    \begin{tikzpicture}
  \def\labelA{1K}
  \def\labelB{10K}
  \def\labelC{20K}

 \begin{axis}[
    width=\linewidth, height=0.6\linewidth,
    xlabel={Time step},
    ylabel={Fidelity},
    xmin=1, xmax=10,
    ymin=0.1, ymax=1.05,
    grid=both,
    grid style={line width=.1pt, draw=gray!30},
    major grid style={line width=.2pt, draw=gray!50},
    legend columns=-1,
    legend cell align=left,
    legend style={
      font=\small,
      draw=gray!50,
      at={(0.5,-0.25)},
      anchor=north ,
      /tikz/every even column/.append style={column sep=0.2cm},
    },
    tick label style={font=\footnotesize},
    label style={font=\small},
    title style={font=\footnotesize},
  ]
    \addplot[gray, dashed, forget plot]
        coordinates {(0,1) (9,1)};
      \addlegendimage{black, thick}                     
      \addlegendentry{shadow mps}                       
      \addlegendimage{black, dashed, thick}                       
      \addlegendentry{direct + mps}                                                                                                               
      \addlegendimage{cbBlue,   mark=*, only marks}               
      \addlegendentry{\labelA}                           
      \addlegendimage{cbPurple, mark=square*, only marks}               
      \addlegendentry{\labelB}                          
      \addlegendimage{cbGreen,  mark=triangle*, only marks}                                                                                               
      \addlegendentry{\labelC}     
      \addplot[cbBlue,   mark=*,    mark options={solid},    thick, forget plot]         table[x index=0, y index=1] {tikz/fidelity_nlat4_ideal.dat};
      \addplot[cbBlue,   mark=*, mark options={solid}, dashed, thick, forget plot] table[x index=0, y index=2] {tikz/fidelity_nlat4_ideal.dat};            
      \addplot[cbPurple, mark=square*,  mark options={solid},      thick, forget plot]         table[x index=0, y index=5] {tikz/fidelity_nlat4_ideal.dat};                                       
      \addplot[cbPurple, mark=square*,  mark options={solid},dashed, thick, forget plot] table[x index=0, y index=6] {tikz/fidelity_nlat4_ideal.dat};                                       
      \addplot[cbGreen,  mark=triangle*,    mark options={solid},    thick, forget plot]         table[x index=0, y index=7] {tikz/fidelity_nlat4_ideal.dat};                                       
      \addplot[cbGreen,  mark=triangle*, mark options={solid}, dashed, thick, forget plot] table[x index=0, y index=8] {tikz/fidelity_nlat4_ideal.dat};                           
    \end{axis}  
\end{tikzpicture}

    \caption{Fidelity $|\bra{\psi}\phi\rangle|^2$ of ideal simulation with different number of shots with shadow MPS (solid lines) with $M=25$ settings and direct state tomography with MPS smoothing (dashed lines) on a $16 \times 16 \times 16$ grid. Shadow MPS outperforms even with a limited budget of 1,000 shots.}
    \label{fig:shadow_shots_ideal}
\end{figure}
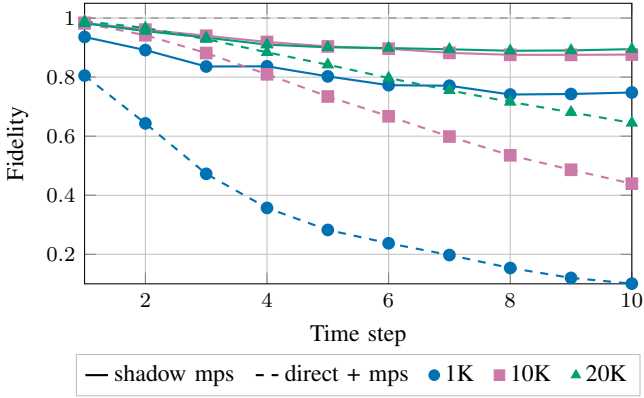

We first compare the performance of the shadow MPS tomography with direct histogram based reconstruction followed by MPS smoothing in an ideal simulation corrupted only with shot noise. Given the same total shot budget (varying between 1,000 to $20{,}000$ shots), Fig.~\ref{fig:shadow_shots_ideal} compares the fidelity of the reconstructed state over ten time steps on a $16^3$ grid. We distribute the shots over $M=25$ settings in the shadow method, while the entire shot budget is used for computational-basis measurement in the direct approach. In the early time steps, with a considerate shot budget of $20{,}000$, both methods perform similarly well. However, the performance with the direct approach with MPS smoothing decays rapidly to a fidelity of $0.64$ at $t=10$ while the shadow MPS is more robust and maintains a fidelity above $0.89$ with $20{,}000$ total shots. Moreover, the gap between the two techniques widens with decreasing shot budget, with the shadow MPS achieving a fidelity of $0.75$ at $t=10$ with $1,000$ shots while the direct approach degrades down to $0.1$ with the same shot budget. 

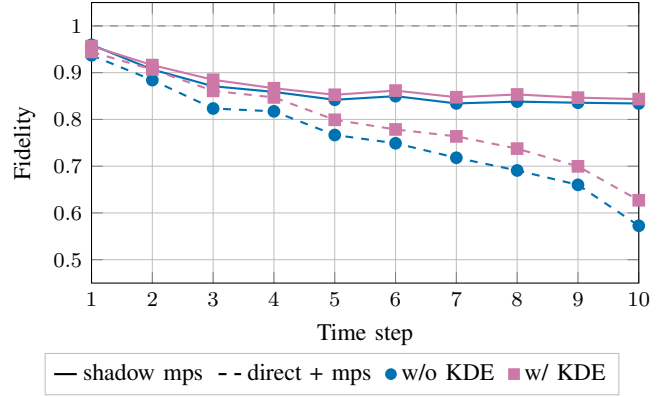
\begin{figure}[!htbp]
    \centering
    \begin{tikzpicture}

 \begin{axis}[
    width=\linewidth, height=0.6\linewidth,
    xlabel={Time step},
    ylabel={Fidelity},
    xmin=1, xmax=10,
    ymin=0.45, ymax=1.05,
    xtick={1,2,3,4,5,6,7,8,9,10},
    ytick={1.0, 0.9, 0.8, 0.7, 0.6, 0.5},
    grid=both,
    grid style={line width=.1pt, draw=gray!30},
    major grid style={line width=.2pt, draw=gray!50},
    legend columns=-1,
    legend cell align=left,
    legend style={
      font=\small,
      draw=gray!50,
      at={(yticklabel cs:-0.4)},  
      anchor=south west,
      /tikz/every even column/.append style={column sep=0.12cm},
    },
    tick label style={font=\footnotesize},
    label style={font=\small},
    title style={font=\footnotesize},
  ]

    \addplot[gray, dashed, forget plot]
      coordinates {(0,1) (9,1)};

      \addlegendimage{black, thick}                     
      \addlegendentry{shadow mps}                       
      \addlegendimage{black, dashed, thick}                       
      \addlegendentry{direct + mps}   

      \addlegendimage{cbBlue, mark=*, only marks}     
      \addlegendentry{w/o KDE}                           
      \addlegendimage{cbPurple, mark=square*, only marks}               
      \addlegendentry{w/ KDE}   
    \addplot[cbBlue,   mark=*, mark options={solid},             thick] table[x index=0, y index=1] {tikz/fidelity_nlat4_qpu.dat};
    
    \addplot[cbPurple,   mark=square*, mark options={solid},                     thick] table[x index=0, y index=2] {tikz/fidelity_nlat4_qpu.dat};

    \addplot[cbBlue, mark=*, mark options={solid},  dashed, thick]     table[x index=0, y index=3] {tikz/fidelity_nlat4_qpu.dat};
     
    \addplot[cbPurple, mark=square*, mark options={solid}, dashed, thick]     table[x index=0, y index=4] {tikz/fidelity_nlat4_qpu.dat};

  \end{axis}
\end{tikzpicture}

    \caption{Fidelity $|\bra{\psi}\phi\rangle|^2$ on IonQ Forte-1 QPU with shadow MPS (solid lines) with $M=25$ settings, 20K shots total vs. direct state tomography with MPS smoothing (dashed lines) on a $16 \times 16 \times 16$ grid. Blue circle and purple squares show results without and with KDE smoothing respectively.}
    \label{fig:shadow_qpu}
\end{figure}

Next, we evaluate both approaches on the IonQ Forte-1 QPU with $20{,}000$ total shots per time step, $M=25$ settings and the results are shown in Fig.~\ref{fig:shadow_qpu}. While fidelity degrades for both techniques on actual hardware relative to the ideal simulation, shadow MPS tomography consistently outperforms the direct approach with a final fidelity of $0.83$ at $t=10$ as compared to $0.57$ for direct tomography. KDE smoothing with bandwidth of $0.5$ is additionally applied to both reconstructions providing a further modest improvement. Due to the LCU post selection step, only about $10\%$ of the shots are actually used in the reconstruction.  The robustness of shadow MPS tomography to both hardware noise and shot noise, together with its ability to capture more information about the state from a limited shot budget makes it well-suited for simulations on finer grids and longer time evolutions.

\begin{figure}[!htbp]
    \centering
    \begin{tikzpicture}

 \begin{axis}[
    width=\linewidth, height=0.6\linewidth,
    xlabel={Time step},
    ylabel={Fidelity},
    xmin=1, xmax=6,
    ymin=0.45, ymax=1.02,
    xtick={1,2,3,4,5,6},
    ytick = {1.0, 0.9, 0.8, 0.7, 0.6, 0.5},
    grid=both,
    grid style={line width=.1pt, draw=gray!30},
    major grid style={line width=.2pt, draw=gray!50},
    legend columns=-1,
    legend cell align=left,
    legend style={
      font=\small,
      draw=gray!50,
      at={(yticklabel cs:-0.4)}, 
      xshift=-2em,
      anchor=south west,
      /tikz/every even column/.append style={column sep=0.15cm},
    },
    tick label style={font=\footnotesize},
    label style={font=\small},
    title style={font=\footnotesize},
  ]

    \addplot[gray, dashed, forget plot]
      coordinates {(0,5) (5,1)};

      \addlegendimage{black, thick}                     
      \addlegendentry{shadow mps}                       
      \addlegendimage{black, dashed, thick}                       
      \addlegendentry{direct+mps}   

      \addlegendimage{cbBlue,mark=*, only marks}     
      \addlegendentry{ideal}                           
      \addlegendimage{cbPurple, mark=square*, only marks}               
      \addlegendentry{noisy}  
                               
      \addlegendimage{cbGreen, mark=triangle*, only marks}               
      \addlegendentry{noisy, w/ KDE}  


    \addplot[cbBlue, mark=*, mark options={solid}, thick] table[x index=0, y index=1] {tikz/fidelity_nlat5.dat};
    \addplot[cbBlue,  mark=*, mark options={solid}, dashed, thick] table[x index=0, y index=2] {tikz/fidelity_nlat5.dat};
    \addplot[cbPurple, mark=square*, mark options={solid}, thick] table[x index=0, y index=3] {tikz/fidelity_nlat5.dat};
    \addplot[cbPurple, mark=square*, mark options={solid},  dashed, thick] table[x index=0, y index=5] {tikz/fidelity_nlat5.dat};
    \addplot[cbGreen, mark=triangle*, mark options={solid},         thick] table[x index=0, y index=4] {tikz/fidelity_nlat5.dat};

    \addplot[cbGreen, mark=triangle*,mark options={solid},   dashed, thick] table[x index=0, y index=6] {tikz/fidelity_nlat5.dat};

  \end{axis}
\end{tikzpicture}

    \caption{Fidelity $|\bra{\psi}\phi\rangle|^2$ on the IonQ Forte-1 noisy simulator with shadow MPS (solid line) with $M=50$ settings, 50K shots total vs. direct state tomography with MPS smoothing (dashed line) on a $32 \times 32 \times 32$ grid. Results with noisy simulator with (green triangles) and without (purple squares) KDE smoothing are shown. Ideal simulation results (blue circles) with the same number of shots are also shown for reference.}
    \label{fig:shadow_noisy_nlat5}
\end{figure}
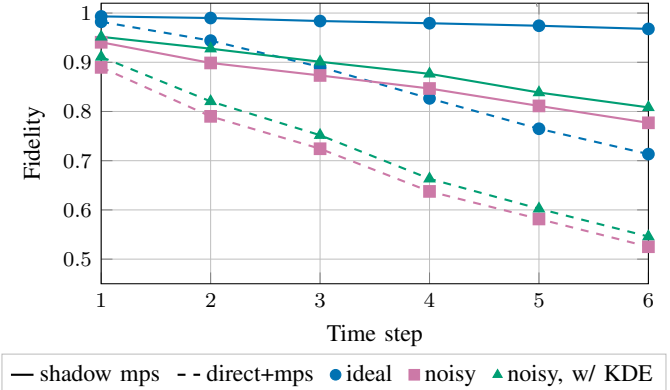

To assess the scalability of the shadow MPS technique to finer grids, we evaluate both approaches on the IonQ noisy simulator with Forte-1 noise model on a $32^3$ grid. The results are shown in Fig.~\ref{fig:shadow_noisy_nlat5} using a total shot budget of $50{,}000$ per time step that is distributed across $M=50$ measurement settings for shadow tomography. There is a significant gap between the two methods that only widens over the course of the simulation. This indicates that the state reconstruction errors compounds more aggressively with the direct approach since the learned state is re-encoded at each time step. While the shadow MPS approach is notably more robust,  more work is needed to lessen the impact of noise and avoid a steep degradation in the quality of reconstruction. 

 \section{General implementation of non-uniform velocity fields} \label{sec:arbitrary_fields}
The hardware results described in this work depend on an efficient construction of quantum operators to implement our spatially varying swirl velocity field. We now show how operators implementing
any divergence-free velocity field can be constructed. 

We will consider the D3Q7 model in this section, for which the lattice weights are $\vec{\omega}=[1/4,1/8,1/8,1/8,1/8,1/8,1/8]$.



We define the following functions:

\begingroup
\allowdisplaybreaks
\begin{align*}
    &g^P_i(\vec r)=\arccos\sqrt{\frac{1+3u_i(\vec r)}{2}},\quad i\in\{x,y,z\}\\
    &g^Q_i(\vec r)=\arccos\sqrt{\frac{1+3u_i(\vec r -\vec \zeta_{i})}{2+3u_i(\vec r -\vec \zeta_{i})-3u_i(\vec r +\vec \zeta_{i})}},\\
    &g^Q_\lambda(\vec r)=\arccos\sqrt{\frac{2+3u_x(\vec r - \vec c_1)-3u_x(\vec r + \vec c_1)}{6}},\\
    &g^Q_\mu(\vec r)=\arccos\sqrt{\frac{2+3u_y(\vec r - \vec c_3)-3u_y(\vec r - \vec c_3)}{4-(3u_x(\vec r - \vec c_5)-3u_x(\vec r + \vec c_5))}}\\
\end{align*}
\endgroup

where $\vec \zeta_x=\vec c_1$, $\vec \zeta_y=\vec c_3$, and $\vec \zeta_z=\vec c_5$.

\begin{algorithm}
\caption{Fast Walsh--Hadamard Transform}
\label{alg:fwht}
\begin{algorithmic}[1]
\Require Array $a$ of length $2^n$, where $n \in \mathbb{Z}^+$
\Ensure $a$ has undergone a Walsh--Hadamard transform
\State $N \gets \mathrm{len}(a)$
\For{$h \gets 1;\; h < N;\; h \gets 2h$}
    \For{$i \gets 0;\; i < N;\; i \gets i+2h$}
        \For{$j \gets i;\; j < i+h;\; j \gets j+1$}
            \State $x \gets a[j];y \gets a[j+h]$
            \State $a[j] \gets y$; $a[j+h] \gets x$
        \EndFor
    \EndFor
    \State $a \gets a/2$
\EndFor
\end{algorithmic}
\end{algorithm}

To implement the operator $U_P$, we first prepare the state

\begin{align}
    |\vec r_0\rangle_G&\left(\frac{1}{2}|0_H\rangle_D+\frac{1}{2}|1_H\rangle_D+\frac{1}{2}|3_H\rangle_D+\frac{1}{2}|5_H\rangle_D\right).
\end{align}
Here, we take the position register to be in a specific basis state $|\vec r_0\rangle$, but it could be in any state when we apply $U_P$.

Following this, we then apply the operation $\sum_{r}|r\rangle_G\langle r|_G\text{RBS}_{D4,D5}(g_x^P(r))$ for the velocity along the $x$ direction, using the so-called RBS gate defined in Ref.~\cite{johri2021nearest},
along with corresponding operations for $y$ and $z$, to get the state in Eq. \eqref{eq:Up}. Similarly we implement $U_Q^\dagger$ by first preparing the state

\begin{equation}
    |\vec r_0\rangle_G\left(\frac{1}{2}|0000001\rangle_D+\frac{\sqrt3}{2}|0000011\rangle_D\right).
\end{equation}

After preparing this state, we apply the operation $\sum_{r}(|r\rangle_G\langle r|_G)\text{RBS}_{D1,D3}(g_\mu^Q(r))\text{RBS}_{D3,D5}(g_\lambda^Q(r))$,
followed by CNOTs on $D0$ from $D1$, $D3$, and $D5$, to retrieve


\begin{align*}
    |\vec r_0\rangle_G\sum_{i\in\{0,1,3,5\}}\frac{\sqrt{2+3u_x(\vec r - \vec c_i)-3u_x(\vec r + \vec c_i)}}{2\sqrt2}|i_H\rangle_D.
\end{align*}

Finally, we apply
\begin{align*}
\sum_{r}&|r\rangle_G\langle r|_G \otimes \prod_{i=0}^3  \left(\text{RBS}^D_{2i,2i+1}(g_i^Q(r))\right)
\end{align*}
with $i=\{0,1,2\}$ = \{z,y,x\}, producing the state in Eq. \eqref{eq:Uq}.

These operations can each be implemented with $N$ multi-controlled RBS gates, or can alternatively use the QPIXL method of Ref.~\cite{amankwah2022quantum} to get a tamer circuit, with $N$ CNOTs and $N$ RBS gates. These QPIXL circuits can be further reduced in size by discarding RBS gates with  angles below a certain threshold. However, regardless of the reduction achieved this way, the clasical processing needed to create the initial QPIXL circuit, before discarding gates, is still $O(N\log(N))$. This scaling ruins our hopes for an end-to-end algorithmic speed-up. For this reason, we propose a faster approximated QPIXL that works for smooth velocity fields.

\begin{algorithm}
\caption{Interpolated 3D FWHT}\label{alg:sparse_fwht_3d}
\begin{algorithmic}[1]
\Require Field $f$, Size $L$, Coarsening $R=2^r$, $K = L/R$
\Ensure Field values on the $L^3$ grid flattened to a vector $v$ of size $N$. $\Theta$ is a dictionary with index-value pairs of the non-zero elements of an approximation of $H^{\otimes n}v$.
\Statex \hrulefill
\State $\mathcal{B} \gets \{ (i,j,k) \cdot R \mid 0 \le i,j,k < K \}$ \Comment{Block base indices}
\State $\mathcal{M} \gets \{ 2^m, L \cdot 2^m, L^2 \cdot 2^m \mid 0 \le m < r \}$ \Comment{Sparse detail offsets}

\For{$b \in \mathcal{B}$}
    \State $\Theta[b] \gets f(\text{midpoint of block } b)$
    \State Compute slopes $\vec{\Delta} = (\Delta_x, \Delta_y, \Delta_z)$ via finite differences with neighbor blocks
    \For{$m \in \{0 \dots r-1\}$}
        \State $\Theta[b + 2^m] \gets \Delta_x 2^{m-1}, \Theta[b + N 2^m] \gets \Delta_y 2^{m-1}, \Theta[b + N^2 2^m] \gets \Delta_z 2^{m-1}$
    \EndFor
\EndFor

\For{$stride \in \{1, L, L^2\}$}
    \State $h \gets R \cdot stride$
    \While{$h < L \cdot stride$}
        \For{$i = 0$ \textbf{to} $N-1$ \textbf{step} $2h$}
            \For{$j \in [i, i+h)$ \textbf{where} $(j \in \mathcal{B} \text{ or } j-base(j) \in \mathcal{M})$}
                \State $x, y \gets \Theta[j], \Theta[j+h]$
                \State $\Theta[j], \Theta[j+h] \gets (x+y)/2, (x-y)/2$
            \EndFor
        \EndFor
        \State $h \gets 2h$
    \EndWhile
\EndFor\\
\Return $\Theta$
\end{algorithmic}
\end{algorithm}

\subsection{Linearly interpolated QPIXL encoding}

\begin{figure}
    \centering
    \includegraphics[width=1\linewidth]{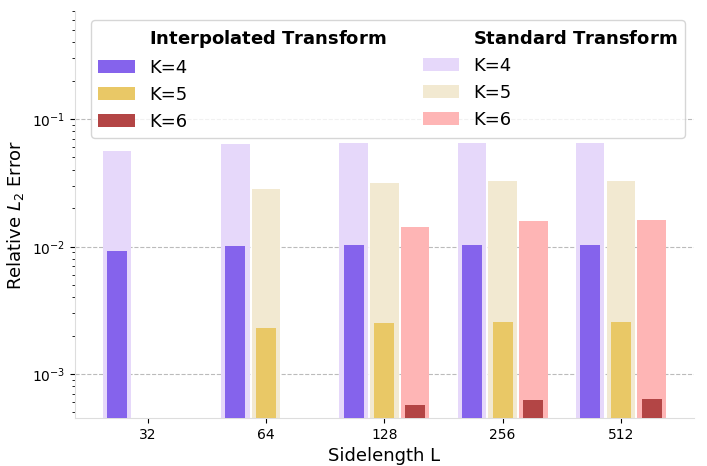}
    \caption{Relative error of the computed Hadamard transform of the flattened $L\times L\times L$ discretization of the field $1+\sin(2\pi x)\sin(2\pi y)\sin(2\pi z)$ for different values of $K$ and $L$. We compare both the interpolated FWHT outlined in Algorithm \ref{alg:sparse_fwht_3d} as well as a coarse version of the standard Hadamard transform that applies $H^{\otimes K}$ since their asymptotic runtimes are comparable.}
\end{figure}

To prepare the input angles for the QPIXL algorithm, we must compute the Walsh-Hadamard Transform (WHT) of a flattened 3D field $\mathcal{F}$. While a standard Fast Walsh-Hadamard Transform (FWHT) requires $O(N \log N)$ operations, we introduce an approximation that reduces complexity to $O(K^3 \log K \log{(N/K)})$, where $K=2^{n-r} < L=2^l$ is the number of sampling points per dimension. This approach assumes the field $\mathcal{F}$ is smooth.

Let us consider a 1-dimensional problem. The standard FWHT algorithm sketched in Algorithm \ref{alg:fwht} applies the operator $I^{\otimes (n-1-i)}\otimes H \otimes I^{\otimes i}$ (up to a scalar factor) at each iteration of the outer while loop. For our modified approach, we split our vector into $K$ blocks each of size $R=2^r$ and approximate the values in each block as a linear function of their element index. The key here is that the Hadamard transform of such a linear block is sparse and easy to compute. The Hadamard transform of a vector $[\theta-2^{r/2}s,\dots,\theta-s,\theta+s,\dots,\theta+2^{r/2}s]^T$ is (up to a constant) given by

\begin{equation*}
    \theta|0\rangle-\sum_{i<r} s2^i|2^i\rangle.
\end{equation*}

Doing this for every block is equivalent to having run the first $r$ iterations of the FWHT algorithm. And when running the remaining $n-r$ iterations of the loop, we only need to consider the $K\log R$ non-zero indices. Thus the overall complexity is $O((n-r)Kr)=O(K\log K \log(N/K))$. The 3-dimensional version of this algorithm is given explicitly in Algorithm \ref{alg:sparse_fwht_3d}.

\section{Simulation of Wall Boundaries}
\label{sec:wall_boundaries}
In this section we explain how we can include objects into the computational domain - whose surfaces are aligned with our lattice grid - such that the passive scalar is barred from diffusing into these objects.

For advection-diffusion, we are already given a classical description of the underlying velocity field, and this velocity field can only be physically valid if the normal velocity at the surface of these walls is zero.

To explain how we can amend our algorithm to block diffusion across walls, we'll start by considering a D2Q5 problem on a square domain with walls on opposing edges. And then we'll extend this for any general wall-boundary in D3Q7.

\subsection{Square domain with opposing walls}

Let's consider a square domain of size $2^l\times2^l$. Our position register will have $2l$ qubits and our direction register will have $5$ qubits.

Let's say we are to add a wall at $y=0$ and $y=2^n-1$. What we want then is to make sure that at $y=1$, there's no streaming in the negative $y$ direction. Similarly, at $y=2^n-2$ there should be no streaming in the positive $y$ direction.

We do this by changing the $k_i$ values at the boundary accordingly. If streaming in the $i$ direction should be barred at a boundary, we will set the new $k_i$ value to be $0$ and increase $k_0$ by the old $k_i$ value, ensuring still that $\sum_ik_i=1$.

\begin{figure}[!t]
    \centering
    \subfloat[$T=0$]{\includegraphics[width=0.24\textwidth]{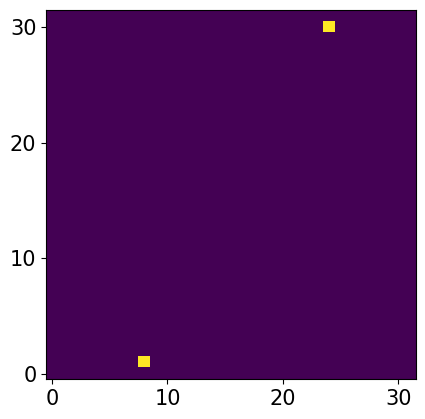}%
    \label{fig_first_case}}
    \hfil
    \subfloat[$T=60$]{\includegraphics[width=0.24\textwidth]{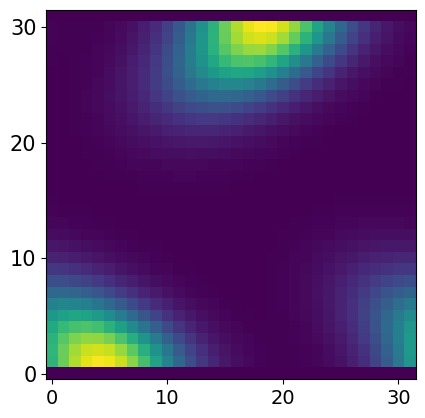}%
    \label{fig_second_case}}
    
    \caption{Simulation of D2Q5 advection-diffusion on a $32\times 32$ grid with velocity field $\vec u(x,y)=(\sin(2\pi y/L)/3,0)$, and walls at $y=0$ and $y=31$}
    \label{fig:walls2d}
\end{figure}

Consider an initial state spanning the $y=1$ line. We start by applying the same $U_P$ operator from Eq. \eqref{eq:Up} to get -

\begin{align*}
    &U_P|r_x\rangle_{x}|1\rangle_{y}|0\rangle_{dir}=
   \ket{r_x}_x\ket{1}_y\bra{r_x}_x\bra{1}_y \left( \sum_{i=0}^4  a_i \ket{i_H}\right),\\
   &\vec{a} = \left[\frac{1}{\sqrt{3}}, \sqrt{\frac{1+3u_x}{6}}, \sqrt{\frac{1-3u_x}{6}},\frac{1}{\sqrt{6}},\frac{1}{\sqrt{6}}\right].
\end{align*}

Here, we will want to revise $k_4$ associated with bitstring $10000$ to be $0$ and we want to increase $k_0$ accordingly. This involves applying a gate that applies a rotation in the subspace of $|00001\rangle$ and $|10000\rangle$ such that the amplitude of the latter is made to be $0$ while the amplitude of the former remains positive.
To do so, we apply an $RBS(\arcsin{\frac{1}{\sqrt{3}}})$ gate between the first and last direction qubits, controlled on the $|1\rangle_y$ state (since we only want this change at the boundary). Applying this operator after $U_P$ 
gives: 
\begin{align*}
    &U_P^\text{new}|r_x\rangle_{x}|1\rangle_{y}|0\rangle_{dir}\\
    =&\left(\text{C}_{|1\rangle_y}\text{RBS}_{d_0,d_4}(\arcsin\frac{1}{\sqrt{3}})U_P\right)|x\rangle_{x}|1\rangle_{y}|0\rangle_{dir}\\
    &=\ket{r_x}\ket{1} \left( \frac{1}{\sqrt{2}}\ket{0_H} + v_x^+ \ket{1_H} + v_x^- \ket{2_H} + \frac{1}{\sqrt{6}} \ket{3_H} \right),
\end{align*}
where $v_x^{\pm} = \sqrt{(1\pm3u_x)/6}$ and, $\text{C}_{|1\rangle_y}\text{RBS}$ is an RBS gate controlled on the $|1\rangle_y$ state.
Similarly, we use an RBS gate before the $U_Q$ operator such that $k_3$ is $0$, i.e., 
\begin{align}
    &U_Q^{\text{new}\dagger}|r_x\rangle_{Gx}|1\rangle_{Gy}|0\rangle_{D}\nonumber\\
    &=\left(\text{C}_{|1\rangle_y}\text{RBS}_{D_0,D_3}(\arcsin{\frac{1}{\sqrt{3}}})U_Q^\dagger\right)|r_x\rangle_{Gx}|1\rangle_{Gy}|0\rangle_{D}\nonumber\\
    &=\ket{r_x}\ket{1} \left( \frac{1}{\sqrt{2}}\ket{0_H} + v_x^+ \ket{1_H} + v_x^- \ket{2_H} + \frac{1}{\sqrt{6}} \ket{4_H} \right).
\end{align}

Similar corrections will need to be applied at $y=2^n-2$. We run simulations on a $32\times32$ grid for $60$ time steps with $\vec u=(\sin(2\pi y/L)/3,0)$. Results are shown in Fig. \ref{fig:walls2d}.

\begin{figure}[!t]
    \centering
    \subfloat[3D view at $T=0$]{\includegraphics[width=0.45\linewidth]{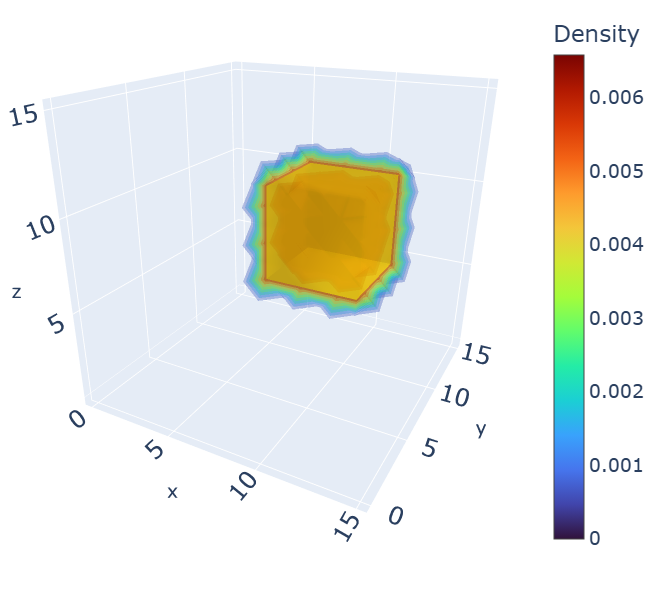}%
    \label{fig_first_case}}
    \hfil
    \subfloat[3D view at $T=24$]{\includegraphics[width=0.45\linewidth]{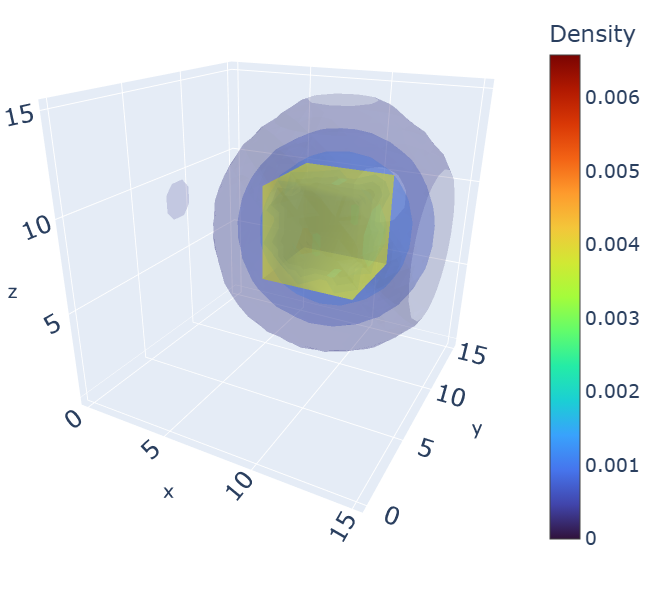}%
    \label{fig_second_case}}
    
    \vspace{-1em} 
    
    \subfloat[Cross section at $y=10$, $T=0$]{\includegraphics[width=0.24\textwidth]{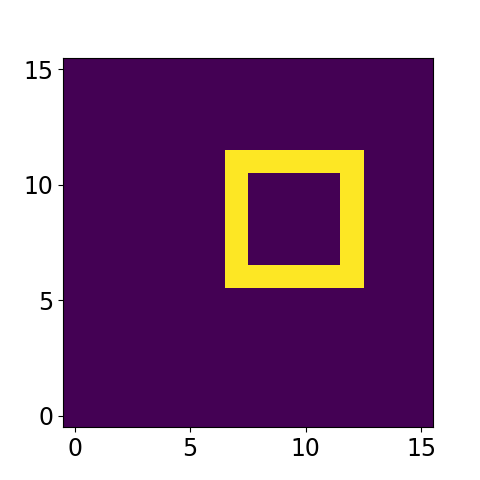}%
    \label{fig_third_case}}
    \hfil
    \subfloat[Cross section at $y=10$, $T=24$]{\includegraphics[width=0.24\textwidth]{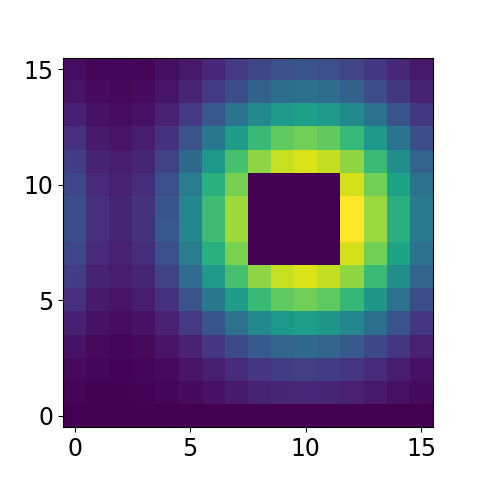}%
    \label{fig_fourth_case}}
    
    \caption{Simulation of species diffusing from the surface of a solid cube suspended in a pipe. The wall oracle $O_W$ defines the pipe by adding walls at $z=0$, $y=0$, and the cube by adding walls at $(x,y,z)\in[8,12]\times[7,11]\times[7,11]$. For the underlying flow, we classically produced a valid divergence-free velocity field such that $u_x(0,y,z)=0.1$.}
    \label{fig:walls3d}
\end{figure}

\subsection{General wall boundaries for D3Q7}

To design a workflow that works for general wall boundaries in D3Q7, we first need an input that specifies what parts of the domain are solid wall. To do this, we consider an oracle $O_W$ that takes in a computational basis state in the position register and flips an auxiliary qubit if that position is a wall and should be inaccessible.

In the last section we showed how to implement wall boundaries by making corrections to the $U_P$ and $U_Q$ operators. We also note that these corrections are not dependent on the velocity field, since they are revisions on $k_i$ and $k_0$, and we know that $k_i(\vec r)=1/8$ when there's a boundary adjacent to $\vec r$ in the $i^\text{th}$ direction (since $u_i(\vec r)=0$ at such a boundary) and that $k_0=1/4$.
Therefore, the operations needed to correct the $U_P$ and $U_Q$ steps only depend on where walls are. Given some specific non-wall position, we only need to know whether there is a wall in each of the $6$ adjacent directions to know what operation to apply to correct the $U_P$ and $U_Q$ operators.

To store this information, we introduce a new wall register (denoted $w$) with $6$ qubits, one for each direction we can move. Given a computational basis state in the position register, we can compute whether there are walls in the $6$ adjacent cells by applying the following operator (with complexity $O(\log{N}+\text{time}(O_W))$):

\begin{equation}
    U_W=\prod_i S_i^{\dagger G} O_W^{G,w_i} S_i^{G}
\end{equation}

When we apply this operator on a state $|\vec r\rangle_{G}|0\rangle_w$, then the $i^\text{th}$ qubit of the wall register will be $1$ iff there is a wall adjacent to $\vec r$ in the $i$ direction.
Each LBM time step is now given by $U_QU_W^\dagger V_Q U_W U_S U_W^\dagger V_P U_W U_P$,
where the $V_P$ and $V_Q$ operators apply appropriate $RBS$ gates in the direction register controlled on the different qubits in the wall register. Since these operators do not involve the grid qubits, their time complexities do not scale with problem size. This means that the overall time complexity needed to take a state $|\Phi_0\rangle$ to a state $|\Phi_T\rangle$ is $$O\left (\frac{\|\Phi_T\|}{\|\Phi_0\|}(K^3\text{polylog}(N)+\text{time}(O_W))\times T\right),$$ where $K$ is the number of points per dimension needed to resolve the velocity field. We use this workflow to run a simulation of species diffusing from the surface of a solid cube suspended in a pipe, depicted in Fig. \ref{fig:walls3d}.

\section{Conclusion and Outlook}
This work advances the quantum lattice Boltzmann method through a hardware demonstration of three‑dimensional scalar transport dynamics under spatially varying velocity fields. We addressed implementation issues of practical relevance, like efficient readout strategies for intermediate data reloading to carry out explicit time-stepping on near term devices. MPS representations combined with kernel density estimation and classical shadow‑based MPS tomography, enable accurate reconstruction of the solution, with reduced measurement overhead and improved robustness to shot noise. 
We further introduced a scalable method for incorporating general velocity fields and wall boundary conditions within the proposed QLBM formalism. Together this enables the full treatment of transport problems involving external and internal flows. 
\par Future work will focus on more comprehensive treatments of boundary conditions, and of bodies with increasingly realistic geometries to study flow fields of direct engineering interest. Beyond individual algorithmic components, we are also interested in  addressing the full CFD workflow, which will require the design of cohesive hybrid quantum–classical routines and well‑defined interfaces between them. Developing such end‑to‑end, workflow‑aware strategies will be critical for evaluating how quantum algorithms may ultimately integrate into industrial‑scale computational physics software.

\section*{Acknowledgement}
The authors would like to thank Masako Yamada for facilitating the logistics of this collaborative project. We also thank Mike Goldman, Ashay Patel, Shantanu Debnath, Ken Wright, and Neal Pisenti for their assistance in running circuits on the Barium experimental testbed.

\bibliographystyle{unsrt}
\bibliography{references}

\end{document}